# Oxygen molecule dissociation on carbon nanostructures with different types of nitrogen doping

Shuang Ni,[a] Zhenyu Li[a] and Jinlong Yang*[a]

5  Energy barrier of oxygen molecule dissociation on carbon nanotube or graphene with different types of nitrogen doping is investigated using density functional theory. The results show that the energy barriers can be reduced efficiently by all types of nitrogen doping in both carbon nanotubes and graphene. Graphite-like nitrogen and Stone-Wales defect nitrogen decrease the energy barrier more efficiently than pyridine-like nitrogen, and a dissociation barrier lower than 0.2 eV can be obtained. Higher nitrogen
10 concentration reduces the energy barrier much more efficiently for graphite-like nitrogen. These observations are closely related to partial occupation of π* orbitals and change of work functions. Our results thus provide useful insights into the oxygen reduction reactions.

## Introduction

Fuel cell provides a clean and efficient method to convert
15 chemical energy into electric energy. Its performance critically depends on the oxygen reduction reactions (ORRs) at the cathode.[1] Traditionally, platinum (Pt) and its alloy nanoparticles are used as electrocatalyst for ORRs. However, the Pt-based electrode is expensive and it suffers from CO poisoning[2] which
20 prohibits the fuel cells from large-scale applications. Therefore, the development of a metal-free ORRs catalyst is very desirable and many efforts have been devoted in this direction.

Recently, Gong et al. reported that metal-free vertically aligned nitrogen-containing carbon nanotubes (VA-NCNTs) has a much
25 better electrocatalytic activity for ORRs compared to the platinum electrodes.[3] Then several studies[4-6] have been performed to get high efficient ORR catalyst based on nitrogen-doped carbon nanotubes (NCNTs), and different synthesis methods to get catalysis NCNTs are also developed. [7-9] Besides NCNTs,
30 other metal-free nitrogen-containing carbon materials, such as nitrogen-doped graphene (N-graphene)[10,11] and nitrogen-doped ordered mesoporous graphitic arrays (NOMGAs),[12] also showed high catalytic activity for ORRs. Even graphitic C3N4 can improve the oxygen reduction activity.[13]

35 Nitrogen atom is expected to play an important role in forming active sites for ORRs, and possible nitrogen forms in carbon nanomaterials have been detected by N 1s XPS spectrum.[14-20] At least three nitrogen forms are found (pyrrole-like, pyridine-like, and graphite-like). Some experiments suggested that pyridine-like
40 nitrogen is responsible for the high ORRs catalysis performance,[21,22] while some other experiments indicated that graphite-like nitrogen is more important.[23,24] Therefore, the most important active sites of nitrogen-doped carbon materials for ORRs catalysis remains unclear.

45 Although there are some theoretical studies[25-29] on oxygen dissociation on nitrogen-doped SWCNT and graphene, a systematic comparison on different types of nitrogen-doping form has still not been reported. In this study, we check energy barrier of oxygen molecule dissociation on carbon nanostructures with
50 different types of nitrogen doping. We find that graphite-like nitrogen is more important than pyridine-like nitrogen for ORRs.

## Computational Details

Calculations were carried out with the density functional theory implemented in the Vienna ab initio simulation package
55 (VASP).[30,31] Perdew, Burke and Ernzerhof (PBE) exchange-correlation functional within the generalized gradient approximation[32] and the projector augmented-wave pseudopotential[33,34] were adopted. We note that the PBE functional generally underestimates the adsorption energy.
60 However, the correction for $O_2$ on (8,0) SWCNT, for example, is about 1.23 kcal/mol.[35] Compare to the energy barrier discussed in this study, it is reasonably small. Spin-polarization was considered in our all calculations.

(8,0) single-wall carbon nanotube (SWCNT) was adopted in
65 our study with the lattice parameter along the tube axis (the $z$ direction ) optimized to be 4.30 Å. The periodic boundary condition was employed, with each SWCNT separated by more than 10 Å vacuum. A cubic 20.0×20.0×12.9 Å$^3$ super cell was used for simulation. All geometry structures were fully relaxed
70 until the force on each atom was less than 0.01 eV/Å. Geometry optimizations were performed with a 1×1×3 Monkhorst-Pack k-point grid,[36] while static calculations were done with a 1×1×5 k-point sampling. The total energies were converged to $10^{-5}$ eV.

Graphene with a 12.78×14.76×15 Å$^3$ supercell was also used.
75 The C-C bond length was set to be 1.42 Å. A 3×3×1 and 5×5×1 k-point grid was adopted for geometry optimizations and static calculations, respectively. The forces convergence is the same as that for (8,0) SWCNT.

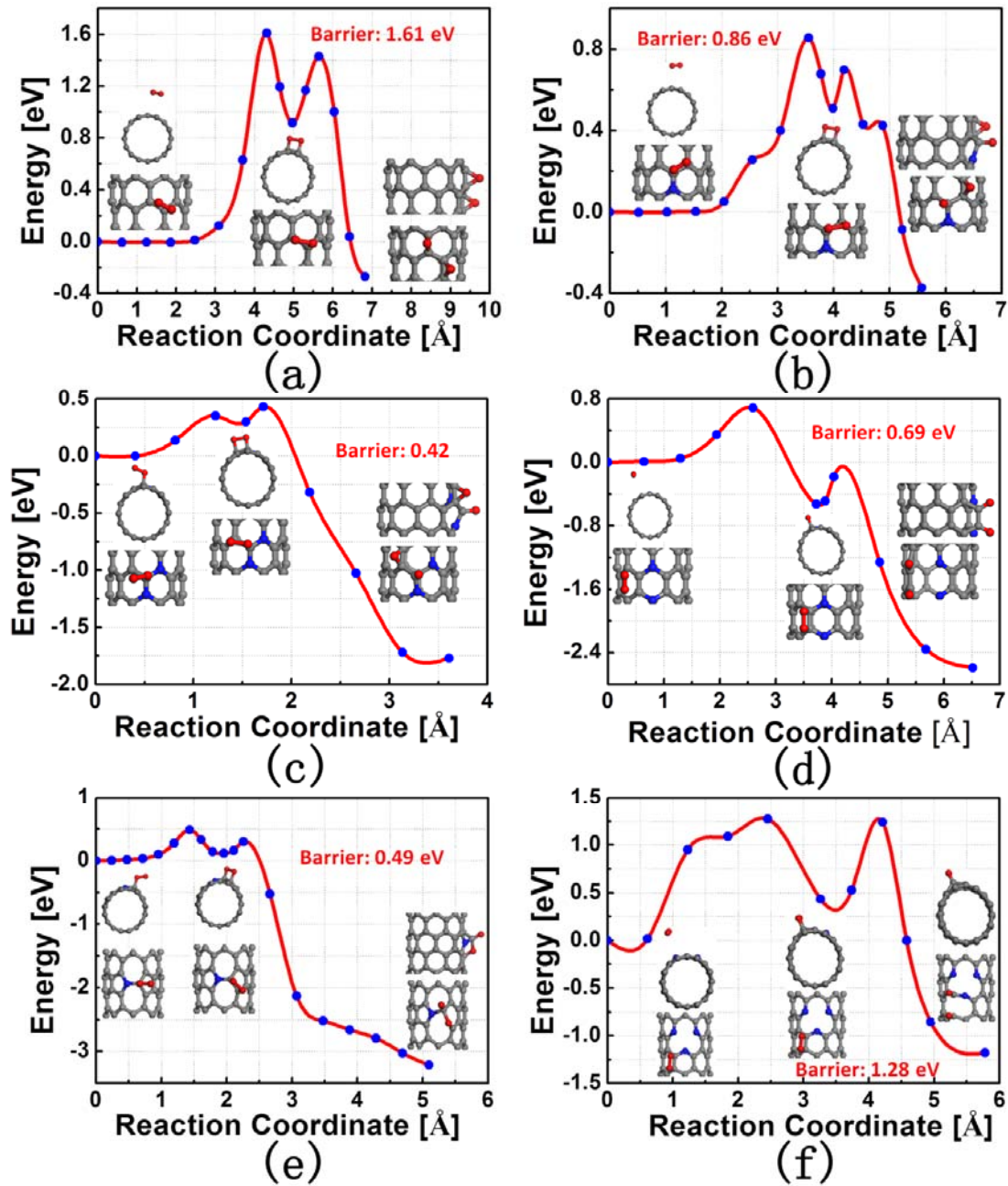

**Fig. 1** Minimal energy paths of oxygen molecule dissociation on (a) a pure (8,0) SWNCT, (b) a one nitrogen atom substituted (8,0) SWCNT, (c) a two *meta* nitrogen atom substituted (8,0) SWCNT, (d) a two *para* nitrogen atom substituted (8,0) SWCNT, (e) a one nitrogen atom substituted (8,0) SWCNT with a stone-wales defect, and (f) a (8,0) SWCNT with pyridine-like nitrogen atoms. Gray dots, blue dots and red dots represent carbon, nitrogen and oxygen atoms respectively.

The climbing image nudged elastic band (CI-NEB) method[37] was used for transition state search. Four to eight images were inserted into initial and final states. The spring force between adjacent images was 5.0 eV/Å. Images were optimized until the forces on each atom are less than 0.02 eV/ Å.

For pyrrole-like nitrogen, we considered the pentagon-ring effect and used one nitrogen atoms substitution in the carbon atom with a stone-wales defect. The pyrridine-like nitrogen was modeled with the Czerw model,[38] where three nitrogen atoms replace three carbon atoms with a shared carbon vacancy. When calculating barrier, we all use the initial state as the energy reference.

## Results

**Pristine SWCNT.** Oxygen molecule dissociation on prinstine SWCNT is considered as a reference point. Initially, we put an oxygen molecule parallel to a C-C bond and 2.3 Å above it. After optimization, the oxygen molecule moves away to 3.3 Å above the bond. The optimized O-O bond length is 1.24 Å, just a little bit longer than that of isolated oxygen molecule. The adsorption energy is very small (0.01 eV). Therefore, the oxygen molecule can be considered as 'floating' on the SWCNTs surface. Then, we consider oxygen molecules approaching the C-C bond and forming a C-O-O-C tetratomic ring. This intermediate state is metastable with a negative adsorption energy. The optimized C-O bond length is 1.48 Å and the O-O bond length increases to 1.51 Å. Finally, we consider this O-O bond broken. Both dissociated oxygen atoms are adsorbed on adjacent C-C bridge sites, which leads to a oxygen-oxygen distance of 3.02 Å.

The energy barriers of these two steps (Fig. 1a) are 1.61 and 1.41 eV, respectively. Our result is similar to the result in ref.39. In their calculation, the energy barrier of singlet $O_2$ dissociation on (8,0) SWCNT is 0.61 eV. Since the triplet $O_2$ is 0.98 eV lower in energy, according to their calculation, the energy barrier for triplet $O_2$ dissociation is 1.59 eV.

**Carbon nanotubes with graphite-like nitrogen.** With a substituted nitrogen atom (Fig. 1b), the oxygen adsorption is still weak. The adsorption energy is 0.04eV, just a little bit larger than that for pristine SWCNT. The optimized O-O bond length is 1.25 Å. For the intermediate state with a C-O-O-C tetratomic ring, the lengths of the two C-O bonds are not same. The one closer to the nitrogen atom is 1.46 Å, while the other is 1.50 Å. The O-O bond length is 1.51 Å. The adsorption energy is also negative, but this structure is more stable than on pristine SWCNT. In order to determine the final dissociated state, we first check single oxygen atom adsorption on the tube. Besides C-C bridge site, it is also possible to adsorb on top of the carbon atoms adjacent to the nitrogen atom. The C-N bridge site is not available for single oxygen atom adsorption.

We consider a final dissociated state with one oxygen atom adsorbed on a carbon top site and the other oxygen atom adsorbed on a C-C bridge site. As shown in Fig. 1b, the energy barriers of the two dissociation steps are 0.86 and 0.70 eV, respectively. Similar result has been obtained by Shan *et al.*[25] for (10,0) SWCNT. Although the energy barrier is much smaller compare to that of pristine SWCNTs. It remains too large for room temperature reaction. We thus further consider increasing the nitrogen-doping concentration.

In Fig. 1c, two nitrogen atoms are placed in *meta* position. The carbon atom adjacent to both nitrogen atoms has more positive charge. Thus, we initially put the oxygen molecule parallel to the C-C bond containing this carbon atom. In the optimized geometry, one oxygen atom bond with this carbon, with a C-O bond length 1.48 Å. The O-O bond length becomes 1.37 Å, longer than the previous two cases. The adsorption energy is about 0.18 eV. Then, in the intermediate state, the two C-O bond lengths are 1.56 and 1.45 Å, respectively. The O-O bond length is 1.51 Å.

We consider the final state with one oxygen atom adsorbed on the carbon top site and the other oxygen atom adsorbed on a C-C bridge site. The adsorption energy of this state is very large, about 1.97eV. One of the C-N bonds breaks with a C-N distance 2.03 Å. The overall energy barrier is only 0.42 eV. Therefore, high concentration of nitrogen doping can significantly reduce the energy barrier, and oxygen dissociation is expected to react under mild conditions.

We have also checked oxygen molecule dissociation on two *para* nitrogen atom substituted (8,0) SWCNT. The total energy barrier is 0.69 eV (fig. 1d), higher than that in the *meta* case.

**Nitrogen-substituted stone-wales defect.** When a nitrogen atom replaces the rotated carbon atom in a pentagon ring of a stone-wales defect (Fig. 1e), the carbon atom in the other pentagon ring adjacent to this nitrogen atom also has positive charge. We initially put the oxygen molecule parallel to a C-C bond which contains this positively charged carbon atom. After optimization, this carbon atom makes a bond with oxygen and the C-O bond length is 1.51 Å. The optimized O-O bond length is 1.33 Å. The adsorption energy of this structure is about 0.49 eV. The intermediate state with both oxygen bonded also has a positive adsorption energy about 0.37 eV, where the O-O bond length is 1.51 Å and the two C-O bond lengths are 1.44 and 1.51 Å. In the final dissociated state, after optimization, one oxygen atom adsorbs on a carbon top site and the other oxygen atom breaks the C-C bond forming a C-O-C ether linkage. The distance of these two carbon atoms is 2.54 Å. The adsorption energy of this state is 3.70 eV. The energy barriers of these two steps are 0.49 and 0.30 eV, respectively (Fig. 1e).

**Carbon nanotubes with pyridine-like nitrogen.** In the initial state (Fig. 1f), the oxygen molecule also floats on SWCNT and the distance between the oxygen molecule and the tube is more than 3.5 Å. The adsorption energy is only 0.01 eV and the O-O bond length is 1.24 Å. In the intermediate state, the O-O bond length is 1.51 Å and two C-O bonds are 1.47 Å in length. The final state is difference from previous one. Both oxygen atoms adsorb on carbon top sites and it breaks the C-C bond. The C-O bond length is about 1.23 Å, indicating carbon-oxygen double bond.

Energy barriers of the two dissociation steps are 1.28 and 1.24 eV, respectively. Although the overall energy barrier is smaller than the pristine SWCNT, it is remain very high.

**Pristine Graphene.** The structures of the initial, medium，and final states are similar to the prinstine SWCNT case (Fig. 2a). The energy barrier of the two steps are 2.62 and 2.71 eV，respectively. Therefore, the total energy barrier is 2.71 eV, even higher than the pristine SWCNT case.

**Graphene with graphite-like nitrogen.** For one nitrogen atom substitution, we get similar initial, intermediate, and final structures (Fig. 2b) compared to the SWCNT case. However, the adsorption energy is very different. For the initial state, the adsorption energy is 0.14 eV, which is much higher than the SWCNT case. Both the intermediate and final states are less stable than the initial state, which leads to an overall energy barrier of 1.87 eV. This is much lower than the pristine graphene case.

We also consider graphene substitution with three nitrogen atoms, following the Okamoto's moldel.[29] In this structure, two nitrogen atoms occupy a *para*-pair position and the third nitrogen sits in the *meta* position with one of them. Oxygen molecule is placed over the C-C bond with its both carbon atoms neighboring to nitrogen atoms (Fig. 2c). The adsorption energy is 0.19 eV. The intermediate state with a C-O-O-C tetratomic ring has an adsorption energy 0.30 eV. The final state has one oxygen atom on carbon top site, and the other oxygen atom on a C-C bridge site. The energy barriers of the two dissociation steps are only 0.19 and 0.02 eV, respectively.

**Graphene with pyridine-like nitrogen.** For graphene with pyridine-like nitrogen, the intermediate state energy is much higher compared to the initial state. This results in a high barrier (1.71 eV) for the first step. The second step further increases the total energy barrier up to 2.34 eV, only somehow lower than the pristine graphene case.

All forms of nitrogen doping lowers barrier of oxygen dissociation on both SWCNT and graphene. The first dissociation step usually determines the overall energy barrier. For N-doped SWCNT with stone-wales defect, one N atom doping already reduces the energy barrier to 0.49 eV, while at least two N atoms are required to reduce energy barrier to such a degree for pristine SWCNT. The pyridine-like nitrogen plays a less important role in reducing oxygen dissociation barrier.

## Discussion

In order to further understanding the nitrogen doping effect, we have studied electronic structure of four different kinds of N-doped SWCNTs (Fig. 1b, Fig. 1c, Fig. 1e and Fig. 1f without oxygen). First, Bader charge analysis[40] has been performed, and the result is marked in Fig. 3. We can see that carbon atoms neighboring to a nitrogen atom are positively charged. Positively charge carbon around nitrogen can be easily understood, based on the relatively higher electronegativity of nitrogen.

Positive charge on carbon atoms around nitrogen has been considered to be a possible reason for the high performance of catalytic activities.[3,41] However, we note that both graphite-like and pyridine-like nitrogen get electron from carbon, but their catalysis abilities are very different. This should be understood from more detailed electronic structures. Graphite-like nitrogen has two $p_z$ electrons. Therefore, $\pi^*$ anti-bonding orbitals around nitrogen will be partially occupied, which may provide efficient active sites. However, for pyridine-like nitrogen, the situation is different. There is a lone pair on nitrogen, and only one electron for the $p_z$ orbital. Occupation in $\pi^*$ anti-bonding orbital is thus expected to be negligible. This may be the reason why pyridine-like nitrogen shows less effective for ORRs.

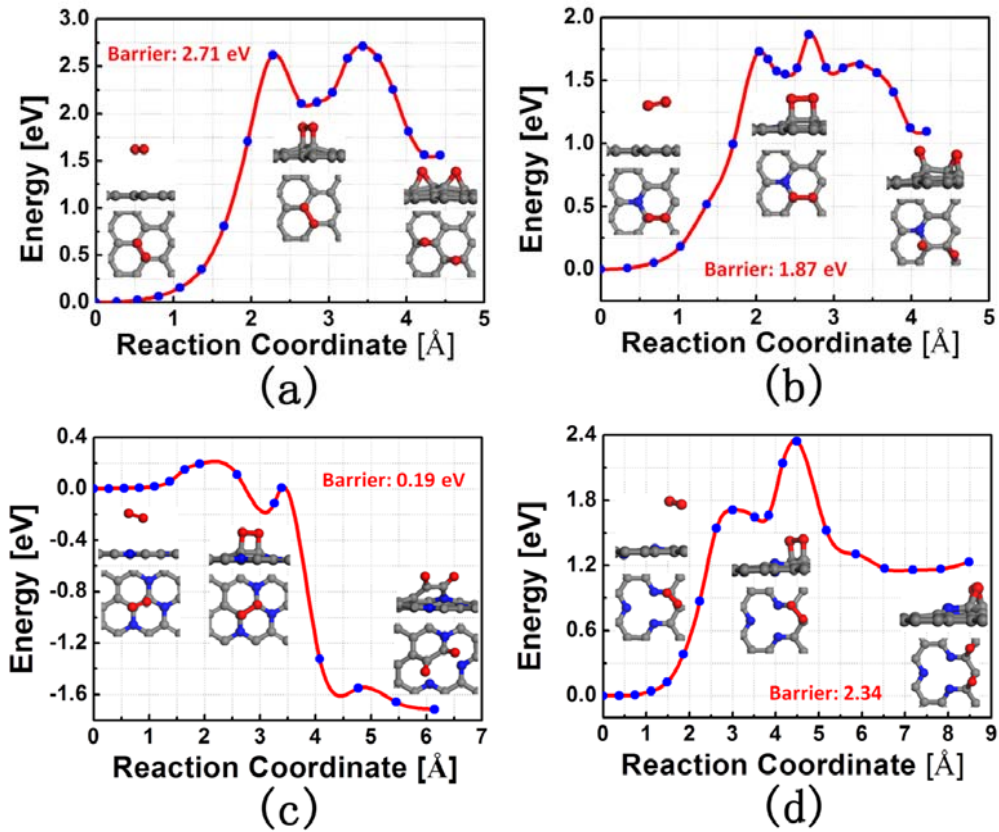

**Fig. 2** Minimal energy paths of oxygen molecules dissociation on (a) pristine graphene, (b) one nitrogen atom substituted graphene, (c) three nitrogen atom substituted graphene, and (d) graphene with pyridine-like nitrogen atoms. Gray dots, blue dots and red dots represent carbon, nitrogen and oxygen atoms respectively.

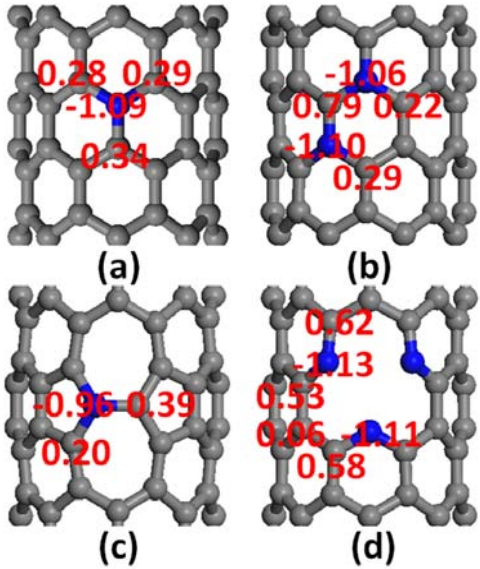

**Fig. 3** Bader charge analysis of (a) one nitrogen atom substituted (8,0) SWCNT, (b) two meta- nitrogen atom substituted (8,0) SWCNT, (c) one nitrogen atom substituted (8,0) SWCNT with a Stone-Wales defect, (d) (8,0) SWCNT with pyridine-like nitrogen. Blue dots and grey dots represent carbon and nitrogen atoms, respectively.

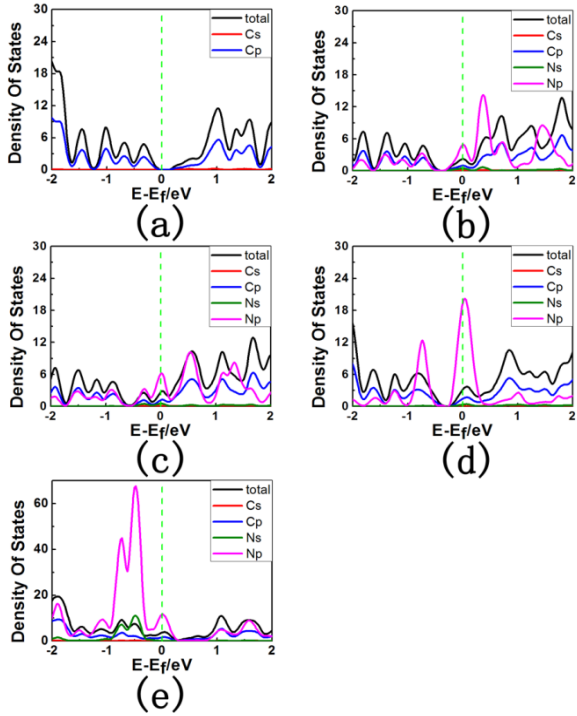

**Fig. 4** Density of states (DOS) of (a) pristine (8,0) SWCNT, (b) one nitrogen atom substituted (8,0) SWCNT, (c) two meta nitrogen atom substituted (8,0) SWCNT, (d) one nitrogen atom substituted (8,0) SWCNT with a Stone-Wales defect, (e) (8,0) SWCNT with pyridine-like nitrogen atoms. The nitrogen DOS is multiplied by the ratio of carbon atom number and nitrogen atom number.

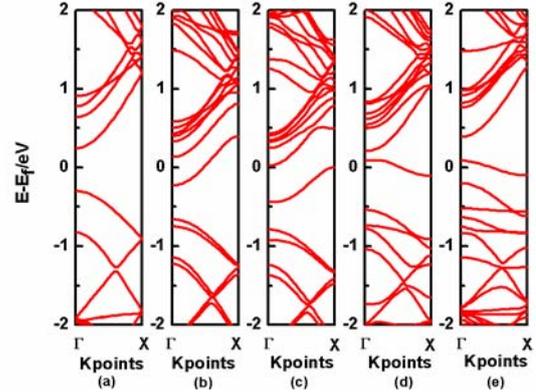

**Fig. 5** Band strutures of (a) pristine (8,0) SWCNT, (b) one nitrogen atom substituted (8,0) SWCNT, (c) two meta nitrogen atom substituted (8,0) SWCNT, (d) one nitrogen atom substitution (8,0) SWCNT with a Stone-Wales defect, (e) (8,0) SWCNT with pyridine-like nitrogen atoms.

Density of states and band structures are plotted in Fig. 4 and Fig. 5, respectively. Pristine (8,0) SWCNT has a gap about 0.54 eV. Based on the relative position of the defect band, we find that doped nitrogen atom with a coordination number of 3 is n-type, while the pyridine-like nitrogen doping is p-type. In Fig. 4e, there are two sharp peaks at the energy -0.50 and -0.76 eV with a nitrogen $p$ character. They are nitrogen lone pair electrons. At the same time, there is a new energy band formed in the energy gap compared to pristine SWCNT (Fig. 5e). This is the defect band due to the pyridine-like nitrogen doping. With a C vacancy among the three nitrogen atoms (Fig. 3d), the system becomes one π electron less, which forms a 'hole' doping. For the other nitrogen-doped SWCNTs, there is no lone pair electron peaks in their DOS. In these system, there is one more π* electron from nitrogen, forming n-type doping. In all doping cases, an impurity band appears in the band gap. This band may directly contribute to the catalysis activity.

Another interesting character of N-doped SWCNT is the upshifted Fermi level, which can be noticed from work functions listed in the Table 1. All kinds of the nitrogen-doing lower work function. Typically, higher energy electrons have stronger reducing power. Therefore, nitrogen-doped SWCNTs show high efficient oxygen reduction ability. The pyridine-like nitrogen doping has the least Fermi energy level shift, thus lower catalysis ability. However, we also note that the high efficient Stone-Wales defect nitrogen-doping doesn't have the very high Fermi energy. In this case, the large adsorption energy of the intermediate state may have made an important role. Based on these analyses, we expect that the pyrrole-like nitrogen may also show a great catalytic activity for ORRs, as it has pentagon ring and a coordination number of there.

**Table 1** Work functions of different nitrogen-doped (8,0) SWCNTs (unit: eV).

| structure | pure (8,0) SWCNT | Fig. 3a | Fig. 3b | Fig. 3c | Fig. 3d |
| --- | --- | --- | --- | --- | --- |
| Work function | 4.84 | 4.22 | 4.14 | 4.55 | 4.69 |

## Conclusion

We have calculated oxygen molecule dissociation on carbon nanotubes and graphene with different types of nitrogen doping. Results show that nitrogen doping can reduce the energy barrier, and graphite-like nitrogen and Stone-Wales defect nitrogen decrease the energy barrier more effectively. Increasing nitrogen concentration may also lower the energy barriers. The main mechanism is based on the work function change and $\pi^*$ and/or impurity band occupation.

## Acknowledgments

This work is partially supported by NSFC (91021004, 20933006, and 20873129), by MOE (FANEDD-2007B23, NCET-08-0521, CUSF), by the National Key Basic Research Program (2011CB921404), by the Fundamental Research Funds for the Central Universities, by USTCSCC, SCCAS, and Shanghai Supercomputer Center.

*a* *Hefei National Laboratory for Physical Sciences at Microscale, University of Science and Technology of China, Hefei,230026 China. E-mail: jlyang@ustc.edu.cn.*